\documentclass[10pt, conference, compsocconf]{IEEEtran}
\usepackage{url}
\usepackage[utf8]{inputenc}
\usepackage{graphicx}
\usepackage{listings}
\usepackage{color}
\usepackage{paralist}
\newcommand{\etal}{{\it et al}.}		
\newcommand{\eg}{{\it{e.g.}}}		
\newcommand{\ie}{{\it{i.e.}}}

\ifCLASSINFOpdf
\else
\fi

\begin{document}
%
\title{Enhanced Usability of Managing Workflows in an Industrial Data Gateway}


\author{\IEEEauthorblockN{Gary A. McGilvary\IEEEauthorrefmark{1},
Malcolm Atkinson\IEEEauthorrefmark{1},
Sandra Gesing\IEEEauthorrefmark{2}, 
Alvaro Aguilera\IEEEauthorrefmark{3},
Richard Grunzke\IEEEauthorrefmark{3} and 
Eva Sciacca\IEEEauthorrefmark{4}}
\IEEEauthorblockA{\IEEEauthorrefmark{1}Edinburgh Data-Intensive Research Group, 
School of Informatics, The University of Edinburgh 
\\ Email: gary.mcgilvary@ed.ac.uk}
\IEEEauthorblockA{\IEEEauthorrefmark{2} Center for Research Computing, University of Notre Dame, Indiana, United States}
\IEEEauthorblockA{\IEEEauthorrefmark{3} Center for Information Services and High Performance Computing (ZIH), Technische Universit\"at Dresden, Germany}
\IEEEauthorblockA{\IEEEauthorrefmark{4} INAF-Osservatorio Astrofisico di Catania, Italy}
}


%


\maketitle

\begin{abstract}
The Grid and Cloud User Support Environment (gUSE) enables users convenient and easy access to grid and cloud infrastructures by providing a general 
purpose, workflow-oriented graphical user interface to create and run workflows on various Distributed Computing Infrastructures (DCIs). 
Its arrangements for creating and modifying existing workflows are, however, non-intuitive and cumbersome due to the technologies and architecture employed by gUSE.
In this paper, we outline the first integrated web-based workflow editor for gUSE with the aim of improving the user experience for those with 
industrial data workflows and the wider gUSE community.
We report initial assessments of the editor's utility based on users' feedback.
We argue that combining access to diverse scalable resources with improved workflow creation tools is important for all big data applications and research infrastructures.

\end{abstract}


%
\IEEEpeerreviewmaketitle

\section{Introduction}
\label{sec:introduction}

A plethora of mature workflow systems has evolved that support diverse workflow concepts and workflow languages with different strengths and focus on different areas of workflow processing. As well as requiring appropriate workflow concepts for their applications, a user community has to evaluate four other requirements:
\begin{inparaenum}[\itshape a\upshape)]
\item	its usability for all members of their community in their work context;
\item	its availability, with respect to licensing terms and cost; 
\item	its anticipated long-term support, \eg\ via an active open-source community; and
\item	its ability to deal efficiently with the scales of data, computation and concurrent use required.
\end{inparaenum} 

The majority of users in the context of the project VAVID \cite{aguilera:2015:a} have no previous exposure to the kind of HPC systems used to power big data analysis. Consequently, the main preference regarding usability has been for a web-based graphical user interface enabling intuitive creation, editing, submission and monitoring of workflows without the need for programming or installations on the users' side. The aspects of scale most critical in the VAVID project are large amounts of data to be processed and a requirement to access high-performance computing infrastructures. The third aspect has been that it should be free of charge also for companies since the project partners are partly from industry. Last but not least, a robust security concept is paramount given the sensitive nature of the industrial data.

gUSE with its flexible web-based user interface WS-PGRADE consists of web services for the workflow management exploiting local clusters as well as
diverse distributed, grid and cloud infrastructures via the ``DCI bridge" \cite{DCI-BridgeChapter2014} \nocite{SCI-BUSbook2014} and 
accessing various distributed data systems via the ``Data Avenue" \cite{DataAvenueSCIbusBookChap5-2014}. 
With these mappings to diverse computing resources and as open source software, gUSE fulfills the requirements for the second, third and fourth criteria for the selection of a workflow system. 
The usability of WS-PGRADE has been found sufficient \nocite{SCI-BUSbook2014} except for the process of creating workflows.

With the WS-PGRADE system prior to the work reported here, users had to create workflows in three stages, one of which required the use a particular graph editor. This editor is a Java Web Start application and therefore requires a local installation of Java and its security preferences to be set correctly; the latter being quite inconvenient for the users particularly within industrial and organizational contexts. For example, conflicts with an organization's security management policies or restrictions on downloads and self-administered installations will often inhibit the use of the gUse workflow editor in such contexts. 

The three-stage creation process also impeded experiment and innovation by requiring completion of one aspect, the topology of data and control flow, for every step of a workflow, before the details of individual steps could be considered. Whereas, a scientist or engineer may want to refine some parts before outlining others, or be able to modify the workflow's graphical representation after a workflow has been created. Both are important in R\&D contexts epitomized by VAVID, as the practitioners need to fluently innovate, refine methods formalized as workflows, incrementally develop workflows and repeatedly use existing workflows on new data or with new parameters---a {{\it modus operandi}} well supported by science gateways \cite{csci284}.

Therefore, the usability issues surrounding gUSE/WS-PGRADE the graph editor, the three-stage workflow creation process and the ability to incrementally develop and refine workflows have been addressed and replaced by the workflow editor presented in this paper. This allows domain scientists to focus and take more responsibility of their own work rather than the technical aspects surrounding it. While this editor is specific to gUSE/WS-PGRADE, it is a step in the direction of also empowering scientists and engineers, improving their prototyping agility and reducing their dependence on IT specialists during innovation \cite{Gesing:2014:WDN:2691175.2691188}. When methods have stabilized and are being used in large scale production the IT specialists may still contribute efficiency and reliability improvements.

The paper establishes the background, and presents the design and implementation of the new editor in that context. An initial evaluation is then reported, that leads to conclusions and plans for further work.

\section{Related Work}
\label{sec:related}
Developers and providers of workflow management systems have recognized the demand  by user communities for usability during the composition of workflows, \ie\ their initial creation and their subsequent edits to improve the method or develop a derived method. 
WS-PGRADE \cite{Kacsuk2012}, Pegasus \cite{Deelman2014}, KNIME \cite{24103053}, Galaxy \cite{Blankenberg2014}, Taverna \cite{Wolstencroft01072013}, Kepler \cite{CPE:CPE994}, Swift \cite{6546066} and UNICORE \cite{DBLP:journals/fgcs/BenedyczakBBMS11} are widely used open-source workflow management systems, which offer workflow canvases. Workflows are illustrated as directed graphs on the canvases. Nodes normally represent jobs or executable modules, while the directed edges define the control and data dependencies between the jobs. 

Conceptually WS-PGRADE distinguishes between an abstract workflow and a concrete workflow. The abstract workflow is created via the graph editor with drag-and-drop mechanisms to add nodes and connect them to each other via input and output ports representing the data flow. The result is a graphical representation of the workflow lacking the information about distinctive jobs or data. In a further step, the abstract workflow is extended to a concrete workflow, which can be configured for concrete jobs, parameters and data files. Similar to gUSE, Pegasus supports a wide range of cluster, grid and cloud infrastructures with cutting-edge data management capabilities. Its web-based user interface is formed by Triana \cite{Taylor:2007wq} but only exists as a prototype.

KNIME follows a different approach to the workflow canvas than WS-PGRADE, that its users find convenient and intuitive. Users select from available modules and nodes that they want to connect with each other. They can develop parts of a workflow completely, including running that subgraph and inspecting intermediate data, before extending the workflow towards completion. 
This allows their focus to match the way they think about a method. 
Advanced users can also create new modules, which requires some programming experience. 

The KNIME workflow canvas is very intuitive but is offered as a workbench based on Eclipse requiring installation on the users' side and not as web-based user interface. This detracts from its utility in contexts such as VAVID.
Galaxy follows a concept for creating workflows similar to the one in KNIME and offers a toolbox via a web-based solution. 

While Galaxy is widely used, especially by the biomedical community, the data management capabilities are quite restricted for large data and necessitate data transfers between single jobs of a workflow to the server hosting the back-end of Galaxy. However, Galaxy can map to the highly parallelized enactments of Swift \cite{6702701}. Another workflow system well established in the biomedical community is Taverna but the workflow canvas is only available as a workbench. The workflows can be shared via the social website myExperiment \cite{DeRoure2009561}.

Kepler offers a desktop application and a web-based graphical user interface for workflow management. The latter has fewer features than the desktop solution and lacks support for creating or modifying a workflow's structure. Thus, it cannot be used for composing a workflow, but only for uploading existing workflows, which can then be modified only with respect to the data and parameters used. While UNICORE also provides both solutions for workflow management and the web-based one is capable of all features available in the desktop application, its use is restricted to computing infrastructures interfaced via UNICORE.

Commercial products offering workflow canvases include a commercial version of KNIME, products applying WS-BPEL (Web Services Business Process Execution Language) \cite{Juric:2006:BPE:1199048}, PipelinePilot \cite{pipelinePilot} or the Genomics Research Platform created by OnRamp \cite{onramp}. The commercial version of KNIME supports advanced features for increasing productivity such as connectors to clouds and Software-as-a-Service (SaaS) as well as features for collaboration. 

WS-BPEL is widely used in industry but requires that all applications integrated into a workflow are available as web service. PipelinePilot, as well the the Genomics Research Platform are solutions that are especially tuned for bioinformatic applications but general applicable for diverse domains. Workflows can be configured for local and batch systems but are missing connectors to grid or cloud infrastructures. Since partners in the VAVID project are from industry, the business models behind such commercial solutions would necessitate the coverage of license costs without delivering more functionalities than gUSE.

In summary, few workflow systems deliver the power of diverse digital resources as gUSE does and most of the web-based creation and editing tools either require local software installations with inherent  security problems or offer incomplete functionality. Hence we suggested a general approach to these deficiencies \cite{Gesing:2014:WDN:2691175.2691188}, however, the current work, though a step in that direction, is specific to gUSE.

\section{Designing the Workflow Editor}
\label{sec:implementation}
In this section, we first give an overview of the pre-existing workflow editing capabilities of gUSE/WS-PGRADE and detail its associated problems. 
We then introduce a partial solution that was under development before discussing the design of our new web-based workflow editor.
We explain how it overcomes the aforementioned inconveniences and how it is integrated into gUSE/WS-PGRADE. 

\subsection{gUSE/WS-PGRADE Graph and Workflow Creation}
\label{sec:guse}
gUSE/WS-PGRADE is composed of a number of Liferay portlets each providing a specific functionality in relation to workflow management. These portlets are typically composed of a presentation layer, portlet layer and persistence layer. The portlet content is displayed using Java Server Pages (JSP), with optional imported JavaScript libraries, where the portlet layer interacts with the client-side presentation layer to serve resources or perform defined actions dependent on the actions of a user. If necessary, the portlet will interact with the database to store or retrieve data.

Using the pre-existing facilities to create a gUSE/WS-PGRADE workflow a user must navigate through three portlets: \textit{Graph}, \textit{Create Concrete} and \textit{Concrete}. A workflow's graph, or an abstract workflow, is created by downloading and executing a Java Network Launch Protocol (JNLP) file from the \textit{Graph} portlet. This instantiates the Java Web Start (JWS) graph editor application, only after the user has correctly added a Java security exception. This process is not user friendly and many problems can arise if the correct security exception is not added or there are problems with the local Java installation. Figure 1 shows an example graph created using the JWS graph editor.

\begin{figure}[ht]
 \begin{center}
\includegraphics[width=0.32\textwidth]{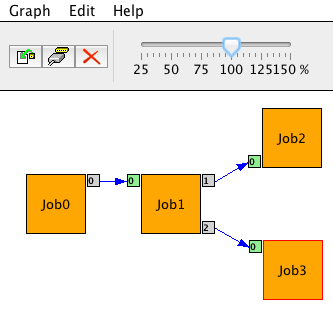}
 \end{center}
 \caption{gUSE Java Web Start Graph Editor}
\end{figure}

\begin{figure*}[ht]
 \begin{center}
\includegraphics[width=0.99\textwidth]{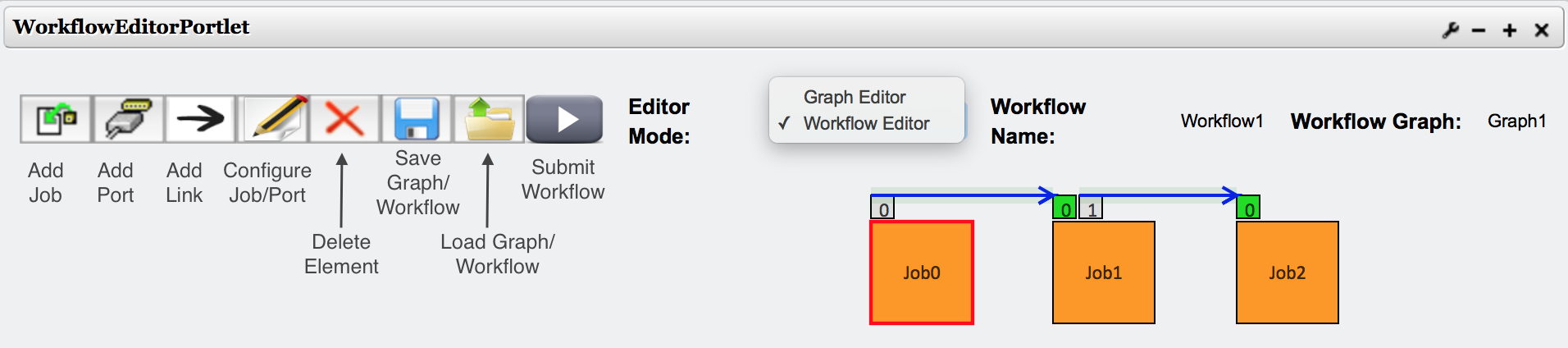}
 \caption{The Web-based gUSE Workflow Editor}
 \end{center}
\end{figure*}

\noindent
Users have the ability to add and remove jobs, input and output ports as well as the connections between ports, all of which are represented as an XML document. After the graph has been saved, it is stored in the gUSE database. 
Graphs can then be transformed into workflows via the \textit{Create Concrete} portlet and configured using the \textit{Concrete} portlet. 
The latter displays a static image of the workflow graph where jobs can be selected allowing configuration parameters to be entered via a pop-up form, \eg\ defining a job's executable type, its arguments and data files. 
Although configuration changes can be made to an existing workflow, the graph's topology and geometry cannot be modified. 
Therefore, when a user wishes to make such changes, a new graph and workflow must be created and re-configured.

\subsection{A Web-based Workflow Editor for gUSE/WS-PGRADE}
\label{sec:wfeditor}

We first introduce a \emph{graph editor} that was being developed contemporaneously, which fed into our design, and then explain the design of our \emph{workflow editor}.

\subsubsection{Graph Editor}
Our workflow editor builds on the previous work of the National Institute of Astrophysics (INAF)\footnote{\url{www.inaf.it/en}} 
that created a web-based graph editor portlet implementation of the JWS graph editor, named \textit{GraphEditorPortlet}. The graph editor was developed in the context of the VisIVO mobile application \cite{vitello2015} to allow gUSE/WS-PGRADE usage from mobile devices, where the JWS editor application cannot operate. The web-based graph editor was developed using the JavaScript libraries KinecticJS 4.7.3\footnote{\url{www.kineticjs.com}}, jQuery 1.9 and jQuery UI 1.10.3\footnote{\url{www.jquery.com}} and replicates the JWS graph editor both in terms of functionality and presentation. Therefore any user familiar with the current JWS graph editor of gUSE/WS-PGRADE will be able to easily use the web-based graph editor. 

The web-based graph editor is split into two components: the graphical editor front-end and the back-end Liferay portlet implementation. Much of the editor's complexity resides with the former, where the position of graphical objects and their respective states must conform to the user's requirements. An object's state consists of the object name, description and its \textit{xy} coordinates. If an object is a port, the port type, its sequence number and a list of any connections to other ports are included. 

The front-end also provides dialogs, similar to those of the JWS graph editor, which must initiate the appropriate operations such as saving and loading graphical representations. Save operations convert each object's state into XML, using the XMLWriter library\footnote{\url{www.javascriptsource.com/ajax/xmlwriter.htm}}, to create an XML document that is passed to the portlet via an AJAX call. The XML is then sent to the gUSE \textit{wfs} module via existing mechanisms to store the graph as an abstract workflow in the gUSE database. Similarly, a load operation retrieves the required graph's XML from \textit{wfs}, which is then passed to KineticJS to reconstruct each object's state on the display canvas.

This web-based graph editor is a direct replacement for the current gUSE JWS graph editor. It does not allow graphs of existing workflows to be modified, nor does it remove the inefficient three-stage process of creating, configuring and submitting workflows.

\subsubsection{Workflow Editor}	
In order to transition from a graph to workflow editor and to solve these usability issues, we developed a new portlet named the \textit{WorkflowEditorPortlet}, which inherits from both the \textit{GraphEditorPortlet} and the \textit{Concrete} portlet but contains additional functionality and improvements to allow the user to directly interact with workflows as opposed to just graphs. The only common entity between the graph and workflow editor is that of the interface and its associated code. Improvements to both the front-end and back-end graph editor components, as well as the necessary additions to gUse, are the foundations of the workflow editor. Figure 2 gives a preview of this complete workflow editor. 

We see that users have the necessary functionality to create, save and load workflows. Furthermore, users have the ability to operate the editor in two modes: \textit{graph} or \textit{workflow}. The former mode is an improved version of the web-based graph editor inherited from INAF, while the latter mode allows direct interactions with workflows, including those created by the JWS graph editor, as well as the ability to submit syntactically correct workflows to a configured DCI. The differentiation of modes ensures past and present users of gUSE/WS-PGRADE are still able to operate on graphs and workflows as individual entities.

In addition to creating this new portlet, we have modified the existing gUSE/WS-PGRADE \textit{Concrete} portlet to exhibit equal functionality to that of the \textit{WorkflowEditorPortlet} by modifying the former's \textit{configure.jsp} presentation layer to include our editor in place of the static workflow image previously provided. In order to ensure both the \textit{WorkflowEditorPortlet} and the \textit{Concrete} portlet provide consistent functionality, both share the same presentation layer, as shown in Figure 3 depicting the editor's architecture.

\begin{figure}[h!]
 \begin{center}
\includegraphics[width=0.5\textwidth]{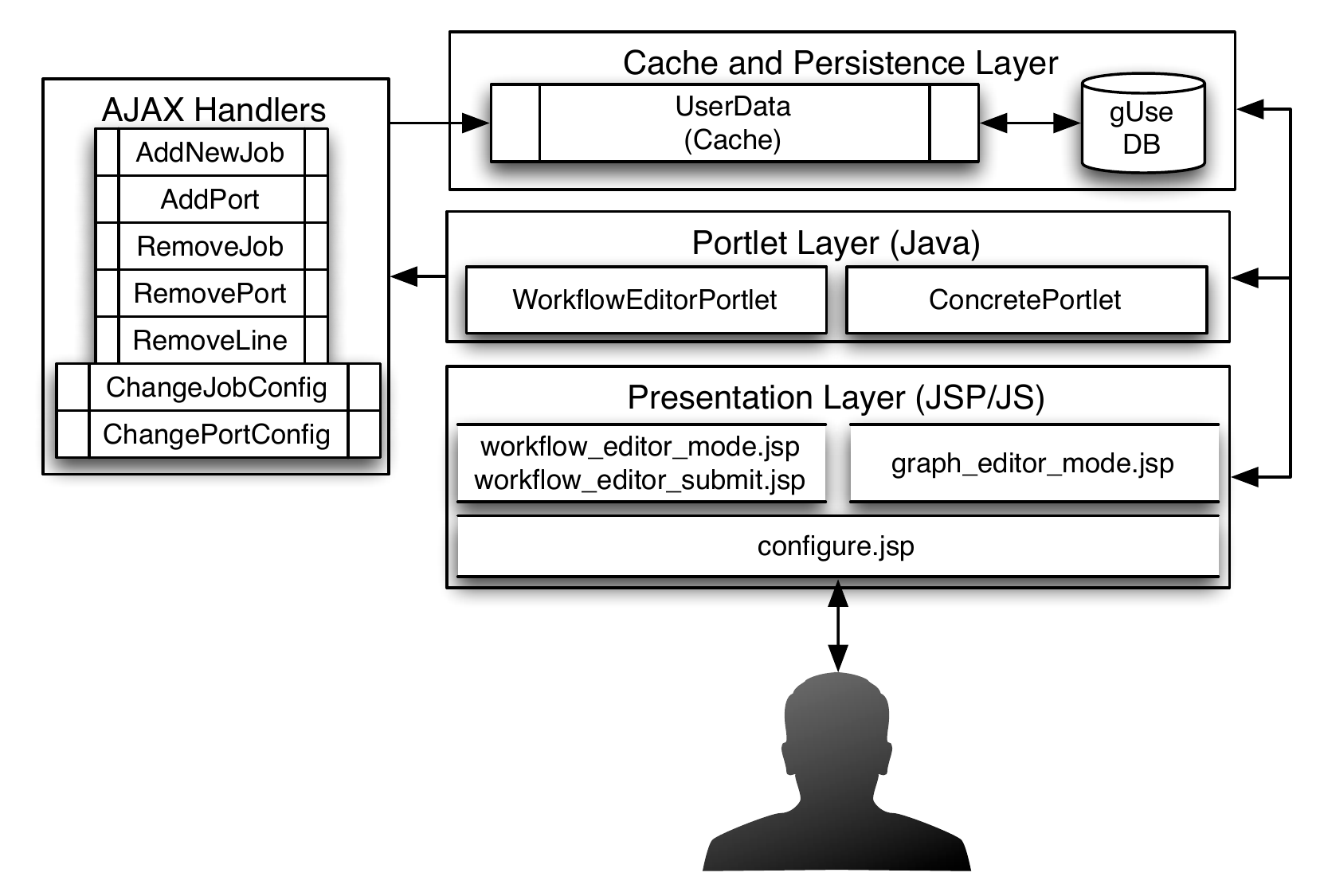}
 \caption{Workflow Editor Architecture}
 \end{center}
\end{figure}
 
In effect, our \textit{WorkflowEditorPortlet} replaces the gUSE/WS-PGRADE \textit{Concrete} portlet, but with added functionality.
The availability of latter remains at the discretion of gUSE. Figure 3 also shows that \textit{configure.jsp} includes the JSP files related to the selected operating mode. Regardless of the mode selected, users continue to interact with the same KineticJS objects, however the integration of workflow editing capabilities required substantial changes to both the graph editor and the gUSE back-end; a task that proved difficult when integrating a solution into a system adopting legacy libraries and where the distinction between front-end and back-end functionality was minimal. 

A large number of these modifications were made to allow graphs of existing workflows to be altered on-demand. 
The previous implementation of gUSE/WS-PGRADE lacks the functionality to save incremental changes to a workflow's graph and instead only permits the bulk saving of graphs and workflows to the database. 
This is a result of storage mechanisms, which cache loaded workflows and only allow configuration parameters to be added or modified. 
Upon a save operation, the cache contents are saved to the database, in turn saving any configuration changes, however any modifications to the graph are not replicated in the cache and therefore are not saved.

We upgraded the cache to account for such changes by creating and instantiating a jQuery AJAX call for each type of change made to the graph. The change is caught and processed by the portlet which is then passed to the appropriate handler to update the cache. This process, as well as the available handlers, are shown in Figure 3.

For example, upon the addition of a new port, the presentation layer concatenates the values of the port’s properties into a string and an AJAX call is made. The portlet processes this call and spawns the \textit{AddPort} handler, which enters the values directly into the cache, either for a new or an existing workflow; the latter resulting in current values being overwritten. The properties of an existing port can be amended via the \textit{ChangePortConfig} handler. The amended cache, present in the Java class \textit{UserData}, can then be stored into the database when a save operation is initiated by the user.

The close conceptual relationship between a gUSE graph and workflow means that in order to allow the user to directly store workflows, it first must be saved as a graph. The workflow can then be created from the graph by calling the existing method \textit{newWorkflow}, which takes the graph name as one of many arguments, and saves the workflow in the database. Similarly, workflows are loaded by determining the graph name of a specified workflow and returning the graph’s XML to reconstruct each object’s state on the display canvas.

The modification of the gUSE cache appears as a trivial addition, however this introduced many complications. Firstly, a new series of database interactions had to be created to retrieve unique identifiers for each new workflow object added to the display canvas. Secondly, any object added to the canvas had to be checked for uniqueness and correctness; a feature that was not present in the inherited web-based graph editor. For example, by adding a port, its name and sequence number must be compared with all others attached to the job.

Validity checks must also ensure objects and their state are consistent with a correctly constructed workflow. For example, validity rules must ensure an output port cannot be connected to another output port. Thirdly, and most importantly, the workflow's state present in the cache must be equivalent to the state present on the display canvas; a feature also not present in the inherited web-based graph editor. If the state is not equivalent in both entities, workflows will be incorrectly configured and subsequently, are likely to exhibit unexpected behaviour when executing on a DCI.

The ability to dynamically add jobs, ports and connections to the cache also allows on-demand workflow configuration. Previously, users had to create and save a workflow before it could be configured via the \textit{Concrete} portlet, by selecting jobs from the static representation of the workflow. By selecting the desired job, users can now instantly add configuration parameters without having to save the workflow in the first instance; all changes are reflected in the cache and are uploaded to the database when the user initiates a save operation.

The incorporation of this feature came with many difficulties, primarily due to the incompatibilities between the different jQuery versions used by the web-based graph editor and the gUSE/WS-PGRADE workflow configuration entry form. The latter uses jQuery 1.3.2 and outdated associated jQuery libraries such as jqDock and BeautyTips. In order to upgrade these libraries, a complete re-design of the gUSE/WS-PGRADE elements reliant on these libraries would have to take place. As the inherited web-based editor is only compatible with jQuery versions 1.9 and above, a solution was devised to operate multiple jQuery versions concurrently. 

The web-based workflow editor provides a much needed solution for the workflow community, and in particular for those who interact with and submit workflows via gUSE. We have shown the necessary changes to create a simple yet effective web-based editor, removing the dependency for a client-side Java installation and extending the Java server portlet implementation. Furthermore, by using standard web technologies, the editor operates on all popular web browsers allowing all users to efficiently create workflows and modify existing ones.

\section{Evaluation}
\label{sec:evaluation}

The new editor and its integrated method for workflow creation and management have been deployed and evaluated on one of the test systems used for the VAVID project; detailed functionality and performance testing of the editor will take place after the use cases of VAVID have been fully created.

When opening the workflow editor portlet, as expected, no Java Web Start application is instantiated and the editor is now displayed inside the web browser. 
As there is no separate editor window, the editor now follows the same style conventions used in the rest of WS-PGRADE.
Furthermore, it is also much faster and involves less user-interaction than downloading and opening the former editor.
The former method was also cumbersome, often involving having to determine how to enable Java support in the web browser and properly adjust the security settings of Java to execute the editor. 

The new editor improves the usability in different scenarios as well.
One of them being the ability to use test systems located behind a remote firewall by simply tunnelling the HTTP port using SSH and accessing the localhost with the browser.
For users without previous exposure to gUSE, the new integrated method of workflow creation, configuration and submission within the same portlet is more intuitive than the previous three-stage method. 
These improvements translate into less helpdesk support required by end users and thus, more time for the development and integration teams to concentrate on other aspects of the VAVID project.


While the general idea of simplifying the three stages of workflow management into a single one is perceived as being more intuitive by the users, the current way of configuring jobs with the new editor could be further improved.
Once the workflow graph is created, users can modify the name and description of each job by double-clicking on it. 
However, selecting any other point of the node that is not its name will display the configuration dialog for the corresponding job.
This behaviour is hinted to the user by highlighting the job's name on mouse over. 
Our experience shows this isn't sufficiently clear for most users independent of their experience level, therefore this will be revised in future versions.

Other potential improvements could be made to the accessibility and positioning of the workflow nodes.
The accessibility problems relate to the color-scheme and style used to render workflows, making certain selections and active elements difficult to recognize.
This is simple to resolve and will be fixed in future releases.
The suboptimal positioning of the elements can be traced back to the JavaScript frameworks upon which the editor is based.
Despite being state of the art when the original INAF-implementation of the editor was created, they have now been superseded by more powerful ones.
Reimplementing the editor with a new framework would have required an effort outside the means of the VAVID project.

An important requirement for the new editor is that of backward compatibility with workflows created using former versions of the editor.
In addition to VAVID's own workflows, the gUSE development team provided a set of test workflows to evaluate the backward compatibility.
No compatibility problems have been found during our tests. 
Previous workflows could be loaded, modified and submitted by the new editor.
Moreover, given that the underlying format in which the workflows are stored in the database hasn't changed, compatibility issues are not expected.
Another vital compatibility aspect is a consistent rendering and functioning of the editor across different browsers and platforms.
During the development and evaluation of the editor, current versions of Mozilla Firefox, Google Chrome, and Safari were used on Linux, OS X, and Microsoft Windows without observing any major changes of the HTML-rendering or a reduction in usability.

Finally, the installation procedure and accompanying documentation of the new editor were also evaluated.
Installing or updating the editor from the source code involves the compilation and re-deployment of the gUSE \textit{frontendbase}, \textit{wfs}, and \textit{wspgrade} modules.
In our experience of using gUSE 3.6.8, this can be performed with little effort by following the installation instructions, if there is a working Java SDK and Apache Maven installed on the system.
It is our hope that the new editor will be integrated into future releases of gUSE making the manual installation unnecessary.

\section{Conclusion and Outlook}
\label{sec:conclusion}
In this paper, we have outlined an improved workflow editor for gUSE/WS-PGRADE that replaces the three-stage process of creating, configuring and submitting workflows, which was unnecessarily cumbersome for prototyping processing and analysis methods and raised conflicts with security policies. 
Our web-based workflow editor portlet implementation directly replaces the gUSE Java Web Start graph editor application and subsequently, the requirement of a local Java installation and correctly specified security preferences. The previous three-stage process of creating workflows has been reduced to a single stage process allowing workflow creation, instant configuration and submission all within our workflow editor portlet.

Furthermore, users now have the ability to dynamically modify the graphical structure of their existing workflows and update job configuration parameters on-demand, allowing the incremental development and refinement of workflows; a feature supported by many other science gateways and a requirement from the users of the VAVID project and many other communities. 

We believe that the aforementioned improvements to the gUSE/WS-PGRADE workflow creation process will greatly enhance the user experience of interacting with workflows allowing domain scientists to focus and take more responsibility of their own work rather than the technical aspects surrounding it.
Preliminary usability studies strongly support this. 
However there are many improvements that could be made to gUSE and to our web-based workflow editor to improve the users' experience and operational behavior further.

The revision of the system's architecture to make the client-side (browser embedded) and server-side of gUSE and WS-PGRADE more independent would be a first step.
The API presented by the server side should support both bulk and incremental changes to workflows. This might be partitioned across several back-end micro-services with
sharply focused functionality to improve flexibility and maintainability \cite{microservices}.
These stable and relevant interfaces would support incremental enhancements to these adopted web-based tools and permit others to create advanced alternatives.

Such workflow editors would exploit novel JavaScript libraries and agile web frameworks. 
For example, the JavaScript library jsPlumb\footnote{\url{www.jsplumb.org}} would improve the visual representation and deliver ready made graphical interaction modes 
because of its excellent design. It offers many features for diverse illustration, representation and manipulation models for the nodes and edges of a workflow graph. 
Also, it is developed by an extensive open-source community, thereby relieving the workflow-editor developers from substantial responsibilities. 

The workflow editor reported here does not use this yet for pragmatic and historical reasons---its adoption is anticipated. 
It underpinned the prototype \emph{generic} workflow editor reported by Gesing \etal\ \cite{Gesing:2014:WDN:2691175.2691188}. 
That proposed web-based workflow editor is intended to accommodate multiple workflow systems for the following reasons:
\begin{inparaenum}[\itshape a\upshape)]
\item	developing powerful and easily learnt web-based GUIs that run on all devices from handhelds to work stations demands skills and effort best amortized over many communities and the similarities between workflow systems make this feasible;
\item	user communities have considerable investments in particular workflow systems that make transfer to replacement workflow systems infeasible, consequently when inter-disciplinary work develops across communities using different systems, and when researchers transfer between groups that consistency saves the researchers intellectual hurdles and delays; and 
\item	the workflow enactment systems are already developing capabilities for integrated multi-workflow language enactments, \eg\ \cite{GKK+13}, and at present developers of the scientific methods have to use each native workflow editor rather than being able to work on the whole method.
\end{inparaenum}

A long-term campaign is required to improve the usability and abstraction so that users who are not adept at computing can nevertheless take full responsibility for the logic of their own methods and can innovate and experiment freely. This becomes ever more necessary as the wealth of available data grows and as more-and-more domain expect to exploit its potential. A broad collaboration across disciplines should address this agenda.

\nocite{workflow-escience}

\section*{Acknowledgment}
The authors would like to thank the Institute for Computer Science and Control (SZTAKI) of the Hungarian Academy of Sciences (MTA) and the gUSE development team for their support throughout this project. The authors would also like to thank the German Federal Ministry of Education and Research (BMBF) for the opportunity to do research in the VAVID project under grant 01IS14005. Furthermore, financial support by the German Research Foundation (DFG) for the MASi project is gratefully acknowledged. 
The research leading to these results has partially been supported by the LSDMA project of the Helmholtz Association of German Research Centres.

\bibliographystyle{IEEEtran}

\bibliography{biblio}

\begin{thebibliography}{10}
\providecommand{\url}[1]{#1}
\csname url@samestyle\endcsname
\providecommand{\newblock}{\relax}
\providecommand{\bibinfo}[2]{#2}
\providecommand{\BIBentrySTDinterwordspacing}{\spaceskip=0pt\relax}
\providecommand{\BIBentryALTinterwordstretchfactor}{4}
\providecommand{\BIBentryALTinterwordspacing}{\spaceskip=\fontdimen2\font plus
\BIBentryALTinterwordstretchfactor\fontdimen3\font minus
  \fontdimen4\font\relax}
\providecommand{\BIBforeignlanguage}[2]{{%
\expandafter\ifx\csname l@#1\endcsname\relax
\typeout{** WARNING: IEEEtran.bst: No hyphenation pattern has been}%
\typeout{** loaded for the language `#1'. Using the pattern for}%
\typeout{** the default language instead.}%
\else
\language=\csname l@#1\endcsname
\fi
#2}}
\providecommand{\BIBdecl}{\relax}
\BIBdecl

\bibitem{aguilera:2015:a}
A.~Aguilera, R.~Grunzke, U.~Markwardt, D.~Habich, D.~Schollbach, and J.~Garcke,
  ``Towards an industry data gateway: An integrated platform for the analysis
  of wind turbine databases,'' in \emph{Science Gateways (IWSG), 2015 7th
  International Workshop on}, accepted.

\bibitem{DCI-BridgeChapter2014}
M.~Kozlovszky, K.~Kar\'{o}czkai, I.~M\'{a}rton, P.~Kacsuk, and T.~Gottdank,
  ``{DCI Bridge: Executing WS-PGRADE Workflows in Distributed Computing
  Infrastructures},'' in \emph{\cite{SCI-BUSbook2014}}, P.~Kacsuk, Ed.\hskip
  1em plus 0.5em minus 0.4em\relax Springer, 2014, ch.~4, pp. 51--67.

\bibitem{SCI-BUSbook2014}
P.~Kacsuk, Ed., \emph{{Science Gateways for Distributed Computing
  Infrastructures: Development framework and exploitation by scientific user
  communities}}.\hskip 1em plus 0.5em minus 0.4em\relax Springer International
  Publishing, 2014.

\bibitem{DataAvenueSCIbusBookChap5-2014}
A.~Hajnal, Z.~Farkas, P.~Kacsuk, and T.~Pint\'{e}r, ``{Remote storage resource
  management in WS-PGRADE/gUSE},'' in \emph{\cite{SCI-BUSbook2014}}, P.~Kacsuk,
  Ed.\hskip 1em plus 0.5em minus 0.4em\relax Springer, 2014, ch.~5, pp. 69--81.

\bibitem{csci284}
A.~Balasko, Z.~Farkas, and P.~Kacsuk, ``{Building science gateways by utilizing
  the generic WS-PGRADE/gUSE workflow system},'' \emph{Computer Science},
  vol.~14, no.~2, 2013.

\bibitem{Gesing:2014:WDN:2691175.2691188}
\BIBentryALTinterwordspacing
S.~Gesing, M.~Atkinson, R.~Filgueira, I.~Taylor, A.~Jones, V.~Stankovski, C.~S.
  Liew, A.~Spinuso, G.~Terstyanszky, and P.~Kacsuk, ``{Workflows in a
  Dashboard: A New Generation of Usability},'' in \emph{Proc. WORKS '14}.\hskip
  1em plus 0.5em minus 0.4em\relax Piscataway, NJ, USA: IEEE Press, 2014, pp.
  82--93. [Online]. Available: \url{http://dx.doi.org/10.1109/WORKS.2014.6}
\BIBentrySTDinterwordspacing

\bibitem{Kacsuk2012}
P.~Kacsuk, Z.~Farkas, M.~Kozlovszky, G.~Hermann, A.~Balasko, K.~Karoczkai, and
  I.~Marton, ``\BIBforeignlanguage{English}{{WS-PGRADE/gUSE Generic DCI Gateway
  Framework for a Large Variety of User Communities}},''
  \emph{\BIBforeignlanguage{English}{Journal of Grid Computing}}, vol.~10,
  no.~4, pp. 601--630, 2012.

\bibitem{Deelman2014}
E.~Deelman, K.~Vahi, G.~Juve, M.~Rynge, S.~Callaghan, P.~J. Maechling,
  R.~Mayani, W.~Chen, R.~F. da~Silva, M.~Livny, and K.~Wenger, ``Pegasus, a
  workflow management system for science automation,'' \emph{Future Gener.
  Comput. Syst.}, no.~0, pp.~--, 2014.

\bibitem{24103053}
S.~Beisken, T.~Meinl, B.~Wiswedel, L.~de~Figueiredo, M.~Berthold, and
  C.~Steinbeck, ``{KNIME-CDK: Workflow-driven cheminformatics},'' \emph{BMC
  Bioinformatics}, vol.~14, no.~1, p. 257, 2013.

\bibitem{Blankenberg2014}
D.~Blankenberg, G.~V. Kuster, N.~Coraor, G.~Ananda, R.~Lazarus, M.~Mangan,
  A.~Nekrutenko, and J.~Taylor, \emph{Galaxy: A Web-Based Genome Analysis Tool
  for Experimentalists}.\hskip 1em plus 0.5em minus 0.4em\relax John Wiley \&
  Sons, Inc., 2010.

\bibitem{Wolstencroft01072013}
K.~Wolstencroft, R.~Haines, D.~Fellows, A.~Williams, D.~Withers, S.~Owen,
  S.~Soiland-Reyes, I.~Dunlop, A.~Nenadic, P.~Fisher, J.~Bhagat, K.~Belhajjame,
  F.~Bacall, A.~Hardisty, A.~Nieva de~la Hidalga, M.~P. Balcazar~Vargas,
  S.~Sufi, and C.~Goble, ``{The Taverna workflow suite: designing and executing
  workflows of Web Services on the desktop, web or in the cloud},''
  \emph{Nucleic Acids Research}, vol.~41, no.~W1, pp. W557--W561, 2013.

\bibitem{CPE:CPE994}
B.~Lud{\"a}scher, I.~Altintas, C.~Berkley, D.~Higgins, E.~Jaeger, M.~Jones,
  E.~A. Lee, J.~Tao, and Y.~Zhao, ``{Scientific workflow management and the
  Kepler system},'' \emph{Concurrency and Computation: Practice and
  Experience}, vol.~18, no.~10, pp. 1039--1065, August 2006.

\bibitem{6546066}
J.~Wozniak, T.~Armstrong, M.~Wilde, D.~Katz, E.~Lusk, and I.~Foster, ``Swift/t:
  Large-scale application composition via distributed-memory dataflow
  processing,'' in \emph{Proc. IEEE/ACM CCGRID '13}, May 2013, pp. 95--102.

\bibitem{DBLP:journals/fgcs/BenedyczakBBMS11}
K.~Benedyczak, P.~Bala, S.~van~den Berghe, R.~Menday, and B.~Schuller, ``Key
  aspects of the {UNICORE} 6 security model,'' \emph{Future Generation Comp.
  Syst.}, vol.~27, no.~2, pp. 195--201, 2011.

\bibitem{Taylor:2007wq}
I.~Taylor, M.~Shields, I.~Wang, and A.~Harrison, ``{The Triana workflow
  environment: Architecture and applications},'' in
  \emph{\cite{workflow-escience}}.\hskip 1em plus 0.5em minus 0.4em\relax
  Springer London, 2007, pp. 320--339.

\bibitem{6702701}
K.~Maheshwari, A.~Rodriguez, D.~Kelly, R.~Madduri, J.~Wozniak, M.~Wilde, and
  I.~Foster, ``{Enabling multi-task computation on Galaxy-based gateways using
  Swift},'' in \emph{CLUSTER 2013}, Sept 2013, pp. 1--3.

\bibitem{DeRoure2009561}
D.~De~Roure, C.~Goble, and R.~Stevens, ``{The design and realisation of the
  myExperiment Virtual Research Environment for social sharing of workflows},''
  \emph{Future Gener. Comput. Syst.}, vol.~25, no.~5, pp. 561--567, 2009.

\bibitem{Juric:2006:BPE:1199048}
M.~B. Juric, \emph{Business Process Execution Language for Web Services BPEL
  and BPEL4WS 2Nd Edition}.\hskip 1em plus 0.5em minus 0.4em\relax Packt
  Publishing, 2006.

\bibitem{pipelinePilot}
\BIBentryALTinterwordspacing
Accelrys, ``Pipeline pilot,'' 2015. [Online]. Available:
  \url{http://accelrys.com/products/collaborative-science/biovia-pipeline-pilot/}
\BIBentrySTDinterwordspacing

\bibitem{onramp}
\BIBentryALTinterwordspacing
OnRamp, ``Genomics research platform,'' 2015. [Online]. Available:
  \url{http://www.onrampbioinformatics.com}
\BIBentrySTDinterwordspacing

\bibitem{vitello2015}
F.~Vitello, E.~Sciacca, U.~Becciani, A.~Costa, P.~Massimino, E.~Takacs, and
  B.~Szakal, ``Mobile application development exploiting science gateway
  technologies,'' \emph{Concurrency and Computation: Practice and Experience},
  2015.

\bibitem{microservices}
M.~Fowler, ``Microservices,''
  \\http://martinfowler.com/articles/microservices.html.

\bibitem{GKK+13}
G.~Terstyanszky, T.~Kukla, T.~Kiss, P.~Kacsuk, A.~Balasko, and Z.~Farkas,
  ``Enabling scientific workflow sharing through coarse-grained
  interoperability,'' \emph{Future Gener. Comput. Syst.}, vol.~37, no.~0, pp.
  46 -- 59, 2014.

\bibitem{workflow-escience}
I.~J. Taylor, E.~Deelman, D.~B. Gannon, and M.~Shields, \emph{Workflows for
  e-Science: Scientific Workflows for Grids}.\hskip 1em plus 0.5em minus
  0.4em\relax Springer London, 2007.

\end{thebibliography}

\end{document}